\documentclass[]{spie}
\usepackage[]{graphicx}
\usepackage{amssymb,amsmath}
\graphicspath{{figures/}}
\usepackage{color}

\title{LASSO: Large Adaptive optics Survey for Substellar Objects using the new SAPHIRA detector on Robo-AO} 

\author{Ma\"issa Salama\supit{a}, James Ou\supit{a}, Christoph Baranec\supit{b}, Michael C. Liu\supit{a}, Brendan P. Bowler\supit{c}, Reed Riddle\supit{d}, Dmitry Duev\supit{d}, Donald Hall\supit{b}, Dani Atkinson\supit{b}, Sean Goebel\supit{b}, Mark Chun\supit{b}, Shane Jacobson\supit{b}, Charles Lockhart\supit{b}, Eric Warmbier\supit{b}, Shrinivas Kulkarni\supit{d}, Nicholas M. Law\supit{e}
\skiplinehalf
\supit{a}Institute for Astronomy, University of Hawai'i at M\=anoa, Honolulu, HI 96822, USA; \\
\supit{b}Institute for Astronomy, University of Hawai'i at M\=anoa, Hilo, HI 96720, USA; \\
\supit{c}Department of Astronomy, The University of Texas at Austin, Austin, TX 78712, USA; \\
\supit{d}Department of Astrophysics, California Institute of Technology, Pasadena, CA 91125, USA; \\
\supit{e}Department of Physics and Astronomy, University of North Carolina at Chapel Hill, Chapel Hill, NC 27599, USA
}

\authorinfo{Further author information: (Send correspondence to M.S.)\\M.S.: E-mail: msalama@ifa.hawaii.edu}

\begin{document} 
\maketitle

\begin{abstract}
We report on initial results from the largest infrared AO direct imaging survey searching for wide orbit ($\gtrsim~100$~AU) massive exoplanets and brown dwarfs as companions around young nearby stars using Robo-AO at the 2.1-m telescope on Kitt Peak, Arizona. The occurrence rates of these rare substellar companions are critical to furthering our understanding of the origin of planetary-mass companions on wide orbits. 
The observing efficiency of Robo-AO allows us to conduct a survey an order of magnitude larger than previously possible. We commissioned a low-noise high-speed SAPHIRA near-infrared camera to conduct this survey and report on its sensitivity, performance, and data reduction process. 
\end{abstract}


\section{INTRODUCTION}
\label{sec:intro} 

Over the past two decades, our knowledge of planetary systems has expanded from just our Solar System to a multitude of planetary architectures. The Kepler mission has detected thousands of exoplanets in close-in orbits ($\lesssim$1~AU) as they transit their host star. Direct imaging identified a complementary population of substellar companions (2--70~M$_{Jup}$), namely planets and brown dwarfs at large projected separations ($\gtrsim$10~AU). The existence of these companions at wide separations has played a critical role in shaping theories about the formation and migration of brown dwarfs and planets, by combining mechanisms such as core accretion, disk instability, cloud fragmentation, and dynamical scattering on various timescales and orbital separations. These mechanisms predict correlations between the presence of a wide-orbit companion and certain environmental characteristics, such as the presence or absence of other companions, circumstellar disk morphologies, and the eccentricity of the companion's orbit.

\begin{figure}
\begin{center}
\includegraphics[width=3.0in]{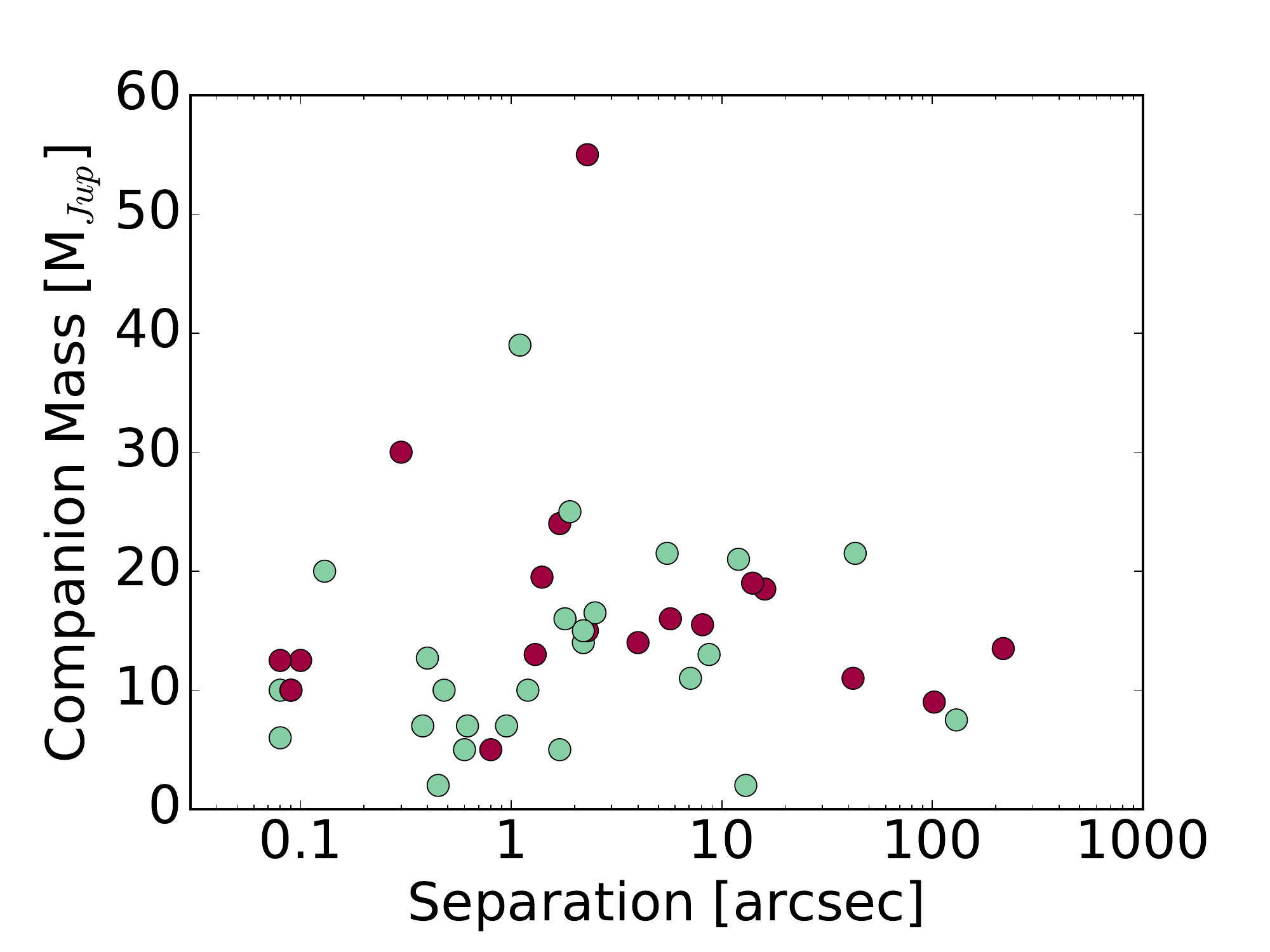}
\caption[fig:known_comps]
{\label{fig:known_comps}Directly imaged substellar companions to date, from Bowler (2016). The burgundy dots are the companions around low-mass ($<$ 0.8~M$_{\odot}$) stars.}
\end{center}
\end{figure}

The IAU currently defines the boundary between brown dwarfs and planets at 13~M$_{Jup}$, but it remains an unresolved issue whether it is an artificial boundary rather than reflective of a natural boundary, linking observational properties of these objects to dominant formation mechanisms (e.g. Chabrier et al. 2014, Schlaufman 2018). Companion demographic studies are necessary to look for trends and clarify the uncertain boundary between brown dwarfs and massive exoplanets. However, due to the small number of discoveries so far, most theoretical work has focused on individual systems. Large exoplanet imaging searches, each on the order of hundreds of targets, have discovered between 0 and 4 substellar companions, bringing the total detections to $\sim$12 objects in the planetary-mass regime ($\lesssim$15~M$_{Jup}$) and another $\sim$30 in the brown dwarf regime (Figure~\ref{fig:known_comps}). To better understand these substellar objects and determine the best parameters for distinguishing between brown dwarfs and massive exoplanets, we need to carry out statistical studies of these objects, which require larger sample sizes. Therefore, the next step is to conduct a survey large enough to greatly boost the detections of these rare objects. This will allow us to go beyond individual discoveries and into population studies to test formation and evolution models. We are conducting the Large Adaptive optics Survey for Substellar Objects (LASSO), the largest infrared direct imaging survey to date, searching for substellar companions around $\sim$3,500 young, low-mass stars in the solar neighborhood. We have conducted a pilot survey, observing $\sim$400 targets using the Robo-AO (Baranec et al. 2014) robotic laser adaptive optics instrument at the Kitt Peak 2.1-m telescope (Salama et al. 2016, Jensen-Clem et al. 2017), equipped with a Selex Avalanche Photodiode for High-speed Infrared Array (SAPHIRA) detector as the infrared science camera. Robo-AO is a robotic laser guide star adaptive optics instrument. Due to the automated operations of Robo-AO, its observing efficiency is unparalleled and enables this survey to be 5$\times$ larger than previously feasible. Robo-AO has been at the 2.1-m telescope on Kitt Peak, Arizona, since November 2015. In June 2018, Robo-AO was decommissioned in preparation for its next commissioning on the UH88 (2.2-m) telescope on Maunakea, by the end of the year 2018.

\section{Robo-AO Infrared Camera}
\label{sec:roboao_irc} 

The signal-to-noise (SNR) ratio can be improved using photon-counting detectors, such as EMCCDs (Electron Multiplying Charge-Coupled Devices) at visible wavelengths, which make use of electron avalanche mechanisms to exponentially multiply the signal without increasing the readnoise. This type of photon-counting technology has recently been extended to infrared wavelengths, with the development of SAPHIRA detectors (Finger et al. 2014). These SAPHIRA detectors allow for almost noiseless signal amplification and ultra-low dark currents (Atkinson et al. 2017), which is especially beneficial for space telescopes where, in the absence of atmospheric interference, detector noise is the main noise source. Such photon-counting IR detectors are also invaluable for ground-based astronomical observations of photon-starved targets. Robo-AO was initially equipped with an EMCCD detector as the visible-light science camera, but in order to extend the scope of observable objects to much cooler and lower-mass objects, namely brown dwarfs and massive exoplanets, we need to move to the near-infrared regime. To this end, we commissioned a new SAPHIRA detector for Robo-AO. This type of high-speed detector is particularly useful to minimize the degrading effect of the target's tip-tilt motion on image quality by taking multiple short-exposure images while adding negligible noise (Jensen-Clem et al. 2017).

Adding an infrared camera allows us to expand the diversity of possible targets and perform multi-color photometry by simultaneously imaging in both the visible and infrared. Additionally, once the infrared camera is fully integrated into the Robo-AO software, it will allow us to point to targets that are otherwise too faint to find in the visible. With the simultaneous infrared and visible imaging we will be able to observe these targets in the optical by increasing their exposure time. Robo-AO does not yet employ active tip-tilt compensation, and instead takes rapid images to subsequently shift and add into a final image. We have previously demonstrated tip-tilt correction with Robo-AO, and will add this to the automated observing routines in the future. We will have the option of performing the tip-tilt correction in the IR, thus expanding our possible sky coverage (Baranec et al. 2015). 


\subsection{SAPHIRA Detector}
\label{sec:SAPHIRA} 

\begin{figure}
\begin{center}
\includegraphics[width=6.5in]{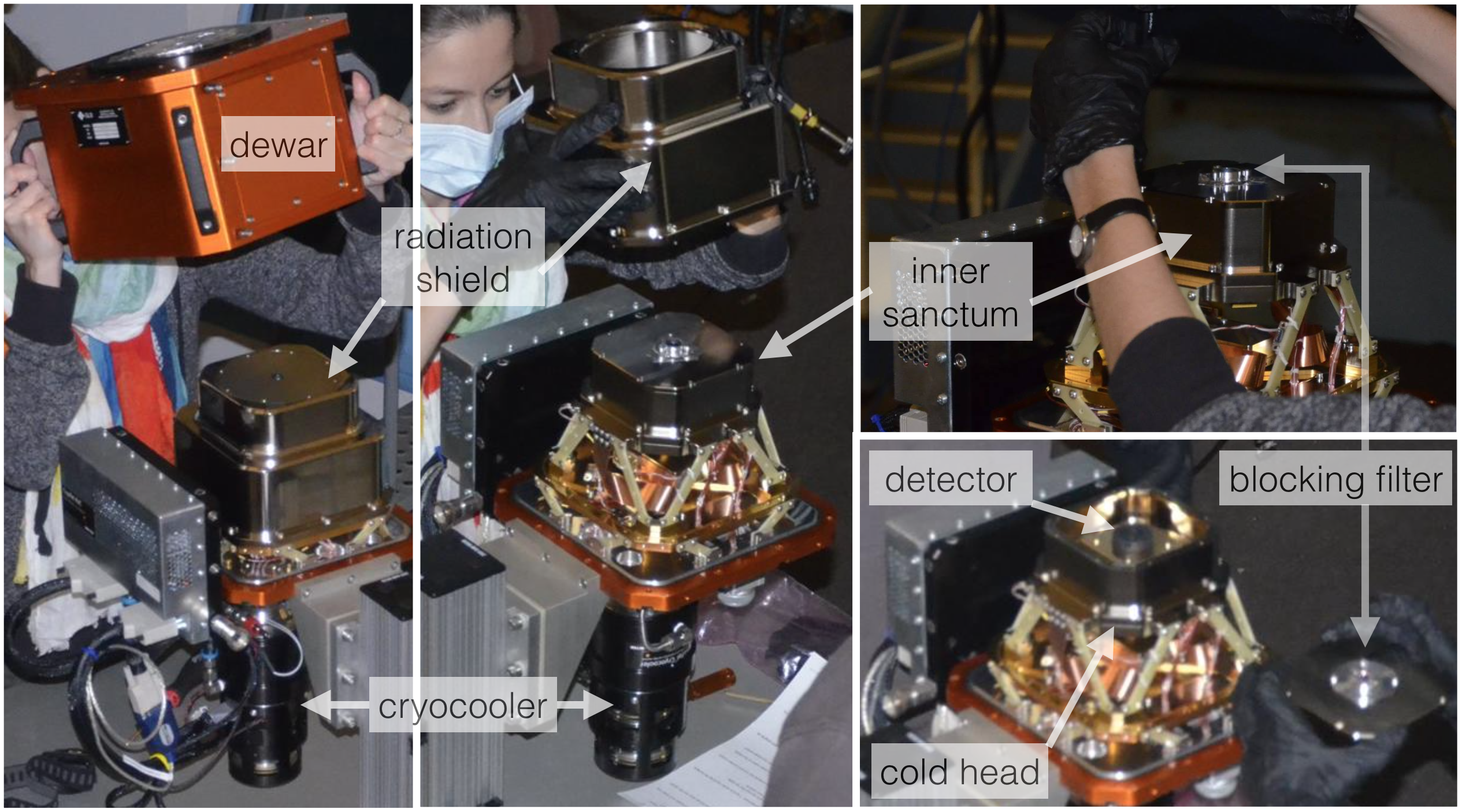}
\caption[fig:camera_pic]
{\label{fig:camera_pic} \textit{Camera interior structure}. From left to right: The dewar (orange) maintains the vacuum seal and encloses the radiation shield, which encloses the inner sanctum. The detector is located inside of the inner sanctum and on top of the cold head. The cold blocking filter is placed on top of the opening of the inner sanctum. The cold head is cryogenically cooled to 65~K and the inner sanctum is heated to 85~K. }
\end{center}
\end{figure}

The SAPHIRA detector (Finger et al. 2014) is a new infrared detector technology that has so far only been used for fringe tracking (Eisenhauer et al. 2016), demonstration tests (Atkinson et al. 2016, Goebel et al. 2016), and with the Robo-AO system (Baranec et al. 2015) at Palomar on a limited basis in 2014. This is the first long-term deployment of such a camera on an adaptive optics system for science applications. Figure~\ref{fig:camera_pic} shows the camera interior setup. The cold head is cryogenically cooled to 65~K. On top of the cold head is the inner sanctum, in which the detector is located. The inner sanctum is heated to 85~K because it is more stable to cool beyond our desired temperature and then heat back up to it. A cold blocking J-filter is placed on the opening at the top of the inner sanctum and a radiation shield surrounds the entire structure to help prevent thermal radiation from reaching the detector. The solid angle open for light to reach the detector is 0.1 steradian. The entire structure is placed inside a vacuum chamber, which is pumped down to $< 1 \times 10^{-5}$ torr in order to isolate the unit from ambient air and maintain its cold temperature. The Institute for Astronomy has developed PB1 (``PizzaBox") electronics to read out the detector. It can be used in different configurations for Hawaii-2RG, Hawaii-4RG, and SAPHIRA detectors. In this case, we use the single board configuration for the $320 \times 256$ SAPHIRA detector, which has 32 readout channels and is capable of reaching a 2~Mpixel/sec sampling rate per channel. This allows for an 800~Hz full frame readout rate. Table~\ref{tab:IR_cam_specs} summarizes the specifications and characteristics of the IR and visible cameras on Robo-AO.

\begin{center}
\begin{tabular}{|l|c|c|}
\hline
& \textbf{Infrared Camera} & \textbf{Visible Camera} \\
\hline
Detector & SAPHIRA & EMCCD \\
Wavelength range &  0.8 -- 2.5 $\mu$m & 400 -- 950 nm \\
Format & 320 $\times$ 256 & 1024 $\times$ 1024 \\
Pixel size & 24 $\mu$m & 13 $\mu$m \\
Field of view & 20.5" $\times$ 16.5" & 36" $\times$ 36"  \\
Plate scale & 0.064 arcsec/pixel & 0.035 arcsec/pixel \\
Filter wheel (cold) & Setup \#1: Short pass & \\
& Setup \#2: J-band & \\
Filter wheel (warm) & Setup \#1: J, H, dark & g', r', i', z', lp600 \\
& Setup \#2: empty &\\
Sampling rate & $\leq$ 800~Hz (using: 20~Hz) & 8.6~Hz \\
Mode & Reset - Read - Read (N times) - Reset & \\
\hline
\end{tabular}
{\label{tab:IR_cam_specs} \\ Table \ref{tab:IR_cam_specs}. Summary of characteristics for Robo-AO science cameras.}
\end{center}

\subsubsection{Automation}
\label{sec:ir_auto} 


Given the short timeframe ($\sim$3~months) before the system was to be removed from the Kitt Peak 2.1-m and overhauled in Hawaii, we decided against integrating control of the SAPHIRA camera with the core software of Robo-AO. Instead, we produced a custom Python script, installed on an independent account on the IR camera control computer with read-only access to the primary control computer. By avoiding direct interaction with the Robo-AO system, we minimized development time and the need for regression testing.

This script automated the IR observation process by monitoring the Robo-AO queue and AO logs.
Observation settings were determined from target entries when they were selected by the queuing
subsystem, with camera activation once the AO loop was closed. For convenience, this script also
triggered initialization of the IR camera system on startup, produced a separate log for IR observations
and comments, and pushed all data to the storage drive on shutdown. As opposed to manual control
of the SAPHIRA camera, this script reduced observation time overhead by approximately 30 seconds
per image (enabling an increase in number of survey observations by up to 15\% in ideal conditions),
and greatly reduced operator training time and errors.

\section{LASSO} 
\label{sec:LASSO}

\subsection{LASSO Target List} 
\label{sec:target_list}

\begin{figure}
\begin{center}
\includegraphics[width=4.5in]{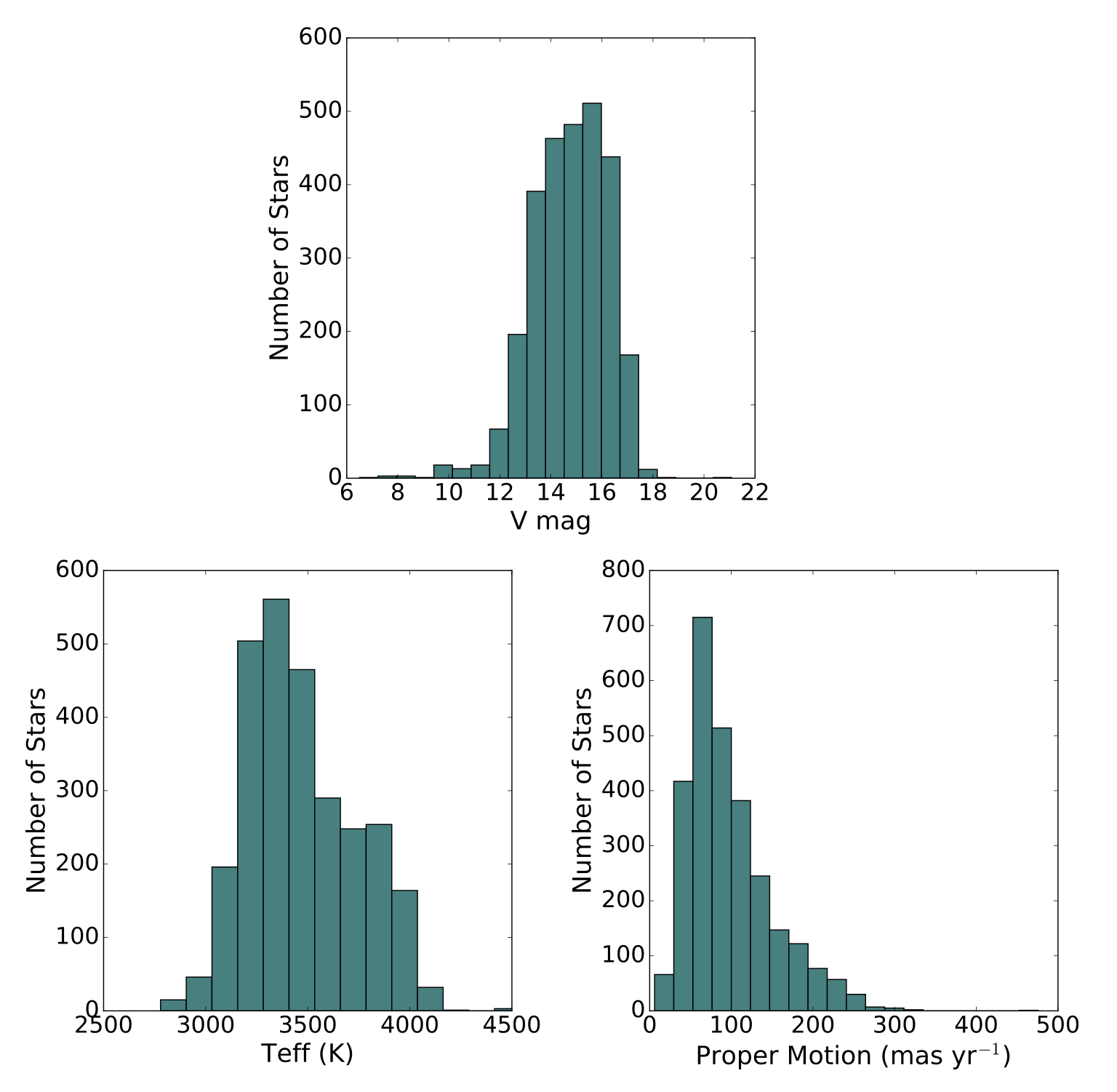}
\caption[fig:target_hists]
{\label{fig:target_hists}Properties of target list constructed from the Cool Dwarf Catalog. The stars are mostly M dwarfs and some late K-type stars. The proper motions are from Gaia DR2. Note the range of V-magnitudes extends beyond the typical 11th magnitude limit for extreme-AO systems due to Robo-AO using a laser guide star system instead of a natural guide star system.}
\end{center}
\end{figure}

As a starting point for building our target list, we used the Cool Dwarf Catalog (CDC), described in Muirhead et al. (2018). Its purpose is to identify cool dwarf targets and their properties for the Transiting Exoplanet Survey Satellite (TESS), which was launched in May 2018. The CDC is incorporated into the TESS Input Catalog and TESS Candidate Target List. The CDC was then cross-matched with the GALEX catalog in order to select for youth, identified by excesses in the UV. The following cuts were applied, following Rodriguez et al. (2013):
\begin{equation}
NUV - W1~(3.4~\mu m) \leq 12.5\ mag
\end{equation}
\begin{equation}
 J - W2~(4.6~\mu m) \geq 0.8\ mag
\end{equation}
\begin{equation}
NUV - W1 < 7 \times (J - W2) + 5.5\ mag
\end{equation}
Next, the remaining targets were cross-matched with Gaia DR2, in order to assemble proper motion and parallax measurements. Finally, observability cuts were applied to ensure that the targets are observable with the Robo-AO system on the Kitt Peak 2.1-m telescope:
\begin{equation}
m_i \leq 15\ mag
\end{equation}
\begin{equation}
Dec > -30^\circ
\end{equation}
Targets with $m_V \leq 17$ mag were selected when no i-band measurement was available.
Finally, only targets with distances, calculated from Gaia DR2 parallaxes, within 100~pc were selected. The resulting list consists of 2,787 stars. The properties of our sample is shown in Figure~\ref{fig:target_hists}. The targets are predominantly M dwarfs with some late K-type stars. They span temperatures of $\sim$3000~--~4000~K, and masses of $\sim$0.1~--~0.8~$M_{\odot}$. Although our contrasts are not as deep as those of extreme-AO systems (since we have no coronagraph and are on smaller, 2-m class, telescopes), we can go beyond the typical primary star brightness limit of 11th V-magnitude because we are using a laser guide star system as opposed to a natural guide star system. The full target list also includes $\sim$1000 additional members of the Sco-Cen association. These stars are at distances slightly beyond 100~pc, but belong to the youngest and nearest association observable form the Northern Hemisphere. The purpose of including these stars in the LASSO survey is to compare young field stars with these younger Sco-Cen members to search for differences in occurrence rates of substellar companions. The total target list consists of $\sim$3500 targets.

\subsection{LASSO Pilot Survey}
\label{sec:performance} 

\begin{figure}
\begin{center}
\includegraphics[width=6.5in]{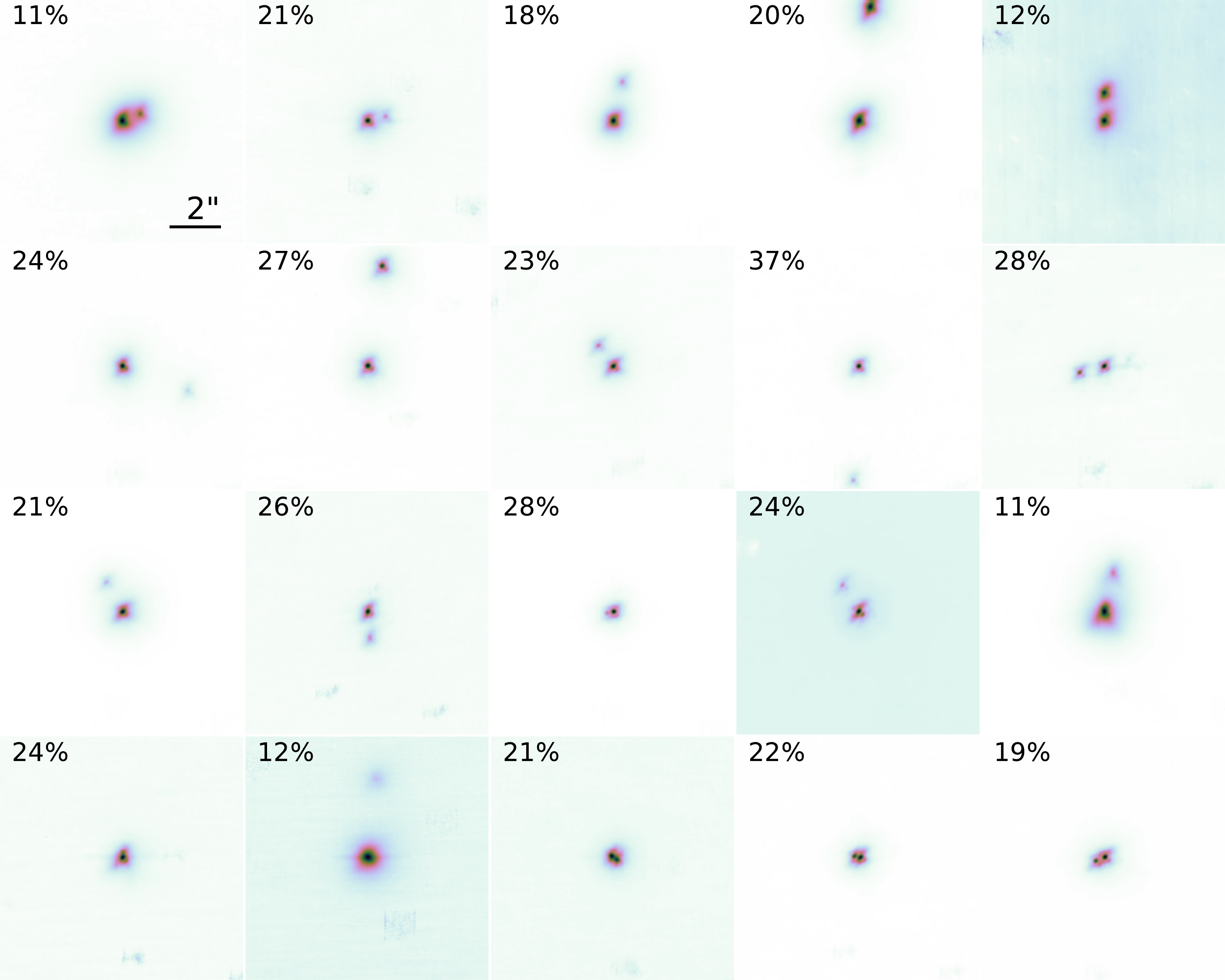}
\caption[fig:image_grid]
{\label{fig:image_grid}Mosaic of example LASSO observations containing companion candidates. The percentage written on each image is the \textit{J}-band Strehl ratio.}
\end{center}
\end{figure}

We have conducted a pilot LASSO survey while Robo-AO was installed on the 2.1-m telescope on Kitt Peak, Arizona. During this time, we observed $\sim$400 stars, through a range of weather and observing conditions, while we were still in the process of testing the SAPHIRA camera readout and automation software. These observations were all carried out in the \textit{J}-band between March and May 2018. Figure~\ref{fig:image_grid} shows a mosaic of LASSO observations with companion candidates. 

\subsubsection{Observing conditions and image quality}

\begin{figure}
\begin{center}
\includegraphics[width=5in]{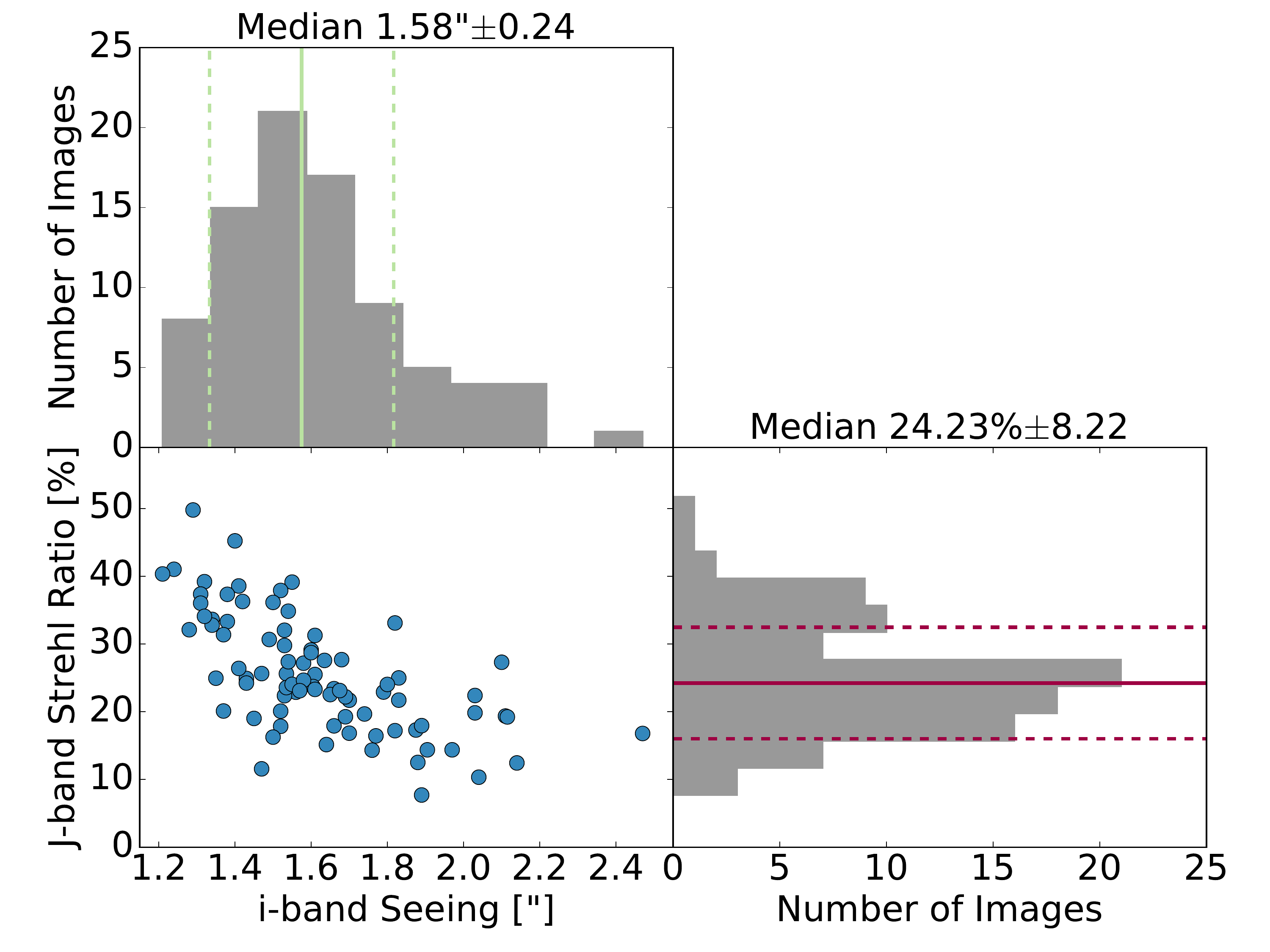}
\caption[fig:seeing_SR]
{\label{fig:seeing_SR} Strehl ratios of pilot LASSO observations from Robo-AO at the 2.1-m telescope on Kitt Peak as a function of seeing conditions. The image Strehl ratios are in the \textit{J}-band and the seeing measurements were made in the \textit{i}'-band. The top plot is a histogram of the seeing measurements and the panel on the right is a histogram of the Strehl ratios. }
\end{center}
\end{figure}

As shown in Figure~\ref{fig:seeing_SR}, during the span of our LASSO observations, the median measured seeing (in the i'-band) was 1.6'' and the median \textit{J}-band Strehl ratio was 24\% (see Salama et al. 2016 for a description of the Strehl ratio calculation). The 2.1-m telescope at Kitt Peak suffers from poor dome seeing, which represent the majority of the contribution to our seeing measurements. The dome also does not have air conditioning, therefore how well the dome and mirror thermalized with the outside air prior to observing has a significant impact on the observing conditions (see Jensen-Clem et al. 2017 for more details).

\subsubsection{Survey sensitivity}

\begin{figure}
\begin{center}
\includegraphics[width=4.5in]{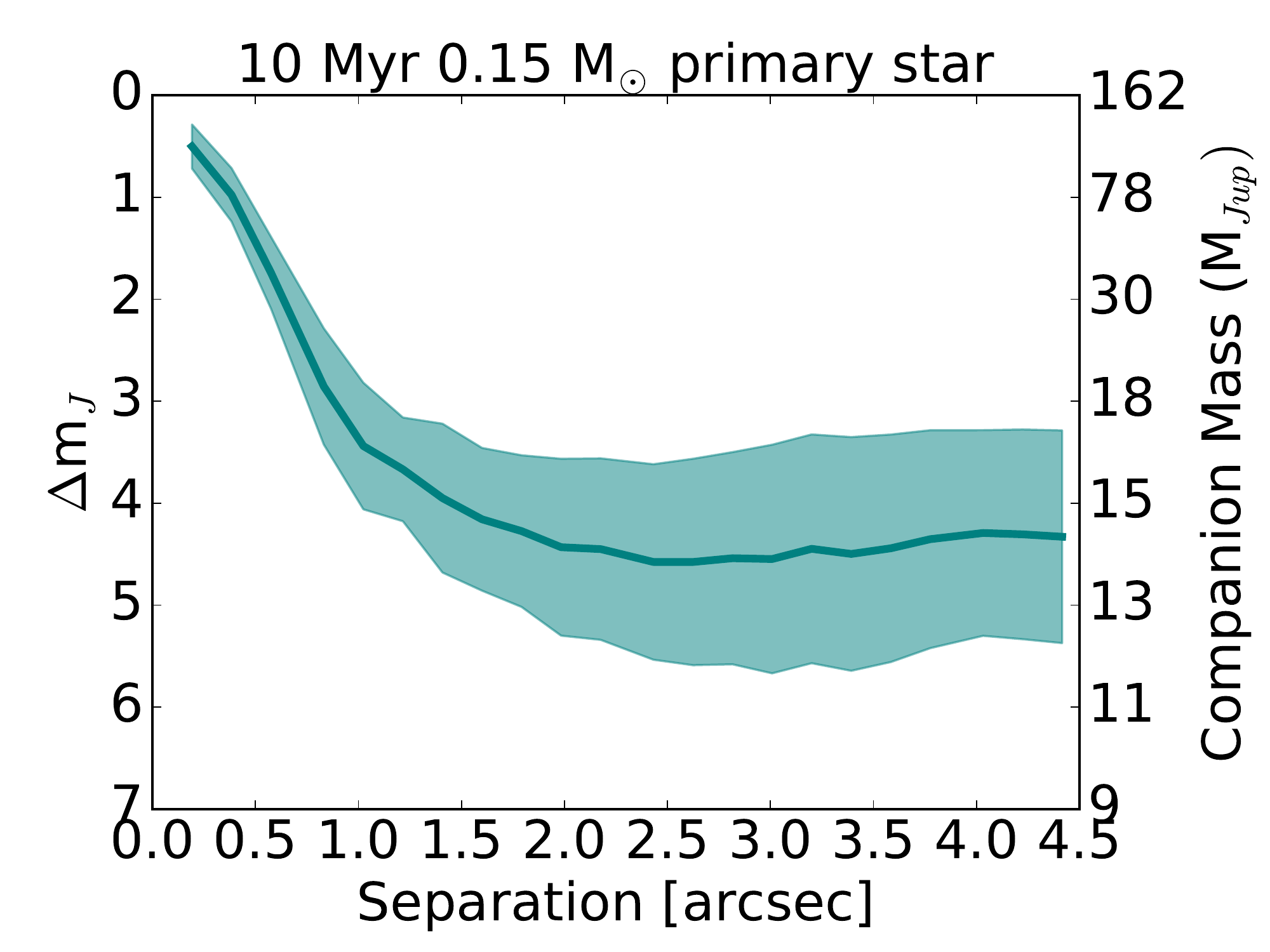}
\caption[fig:contrast_mcomp]
{\label{fig:contrast_mcomp}Median 5-$\sigma$ contrast curve and one standard deviation of pilot LASSO observations from Robo-AO at the 2.1-m telescope on Kitt Peak. The y-axis on the right shows what the contrasts correspond to, in terms of companion masses, if we assume our most sensitive case of a primary star of 0.15~$M_{\odot}$ and 10~Myrs old.}
\end{center}
\end{figure}

\begin{figure}
\begin{center}
\includegraphics[width=4.5in]{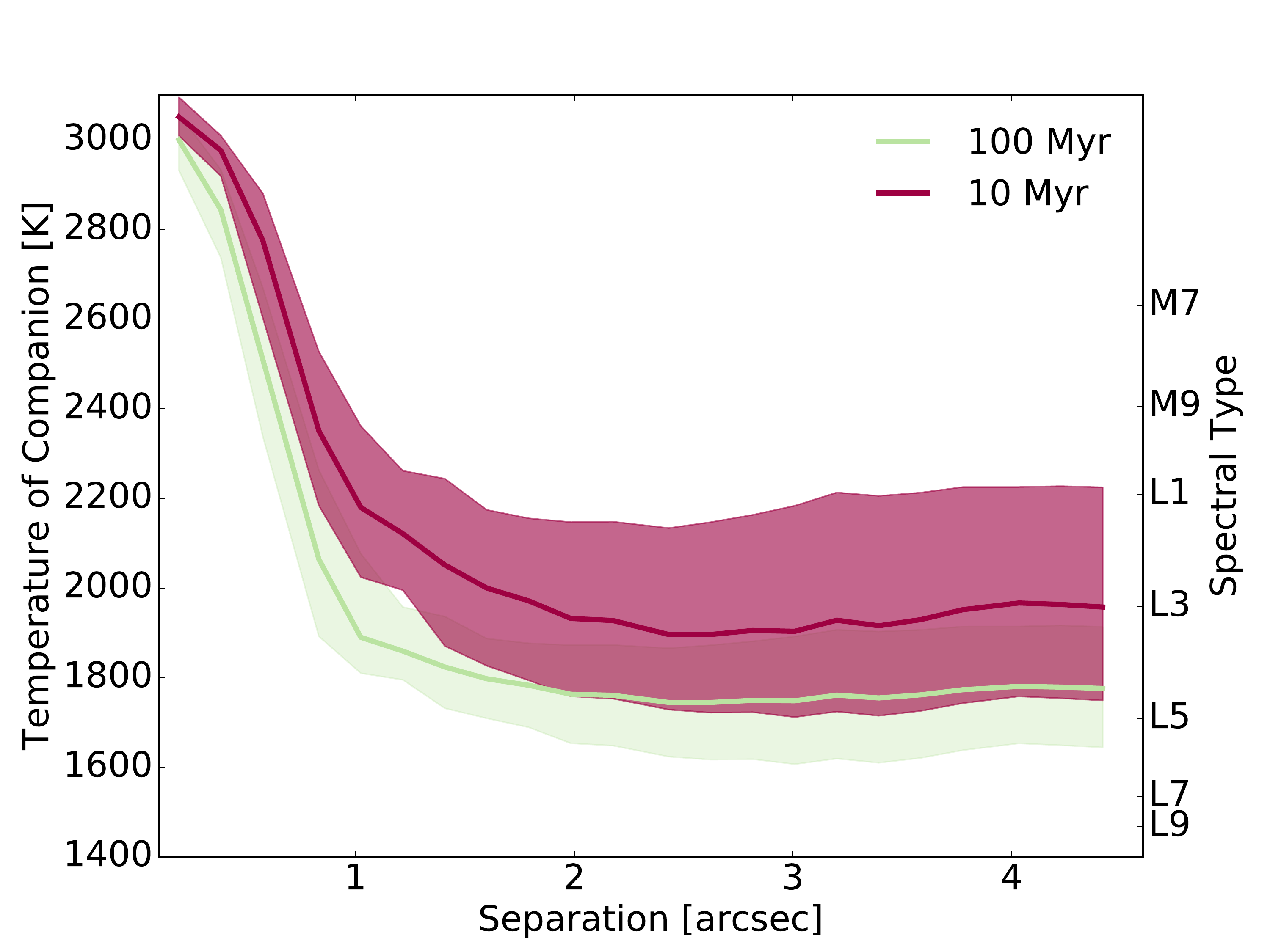}
\caption[fig:contrast_tcomp]
{\label{fig:contrast_tcomp}Same median 5-$\sigma$ contrast curve as Figure \ref{fig:contrast_mcomp} of pilot LASSO observations from Robo-AO at the 2.1-m telescope on Kitt Peak but in terms of companion temperatures and spectral types and shown for the 10~Myr and 100~Myr cases. The primary is again assumed to be 0.15~$M_{\odot}$.}
\end{center}
\end{figure}

To assess the sensitivity of the LASSO survey, specifically with the goal of detecting faint substellar companions, we injected fake companions into each real image and traced the limiting magnitude contrast and separation at which we could still recover the companion. We injected an artificial companion by taking the star's PSF and scaling it down to a range of contrasts, then shifting over a range of projected separations. A radial average is subtracted from the image with the injected companion to suppress the central star, then this image is divided by the root mean square noise of each radial ring, thus creating a Signal-to-Noise ratio (SNR) map. A minimum SNR of 5 was our threshold for declaring that the injected companion was recovered. Figure~\ref{fig:contrast_mcomp} shows the resulting median contrast curve as well as the 1-sigma error bars (shaded region).
If we assume the most sensitive case of a primary star of 0.15~$M_{\odot}$ that is 10~Myrs old, then the contrasts shown in Figure~\ref{fig:contrast_mcomp} correspond to the companion masses displayed on the right y-axis. We converted the magnitude of the companion to a mass using hot-start evolutionary models (Chabrier et al. 2000, Baraffe et al. 2015). Figure~\ref{fig:contrast_tcomp} shows the same median contrast curve but in terms of companion temperatures and spectral classes and for the same assumed primary stars, except at 10~Myrs and 100~Myrs. 

\section{Conclusion}
\label{sec:Concl}

We are conducting the LASSO survey using the Robo-AO instrument equipped with a SAPHIRA detector. The purpose of this program is to search for massive exoplanets and brown dwarfs as wide-orbit companions to young, nearby, low-mass stars. Robo-AO enables us to conduct this large direct imaging survey of $\sim$3500 targets, which is 5--10 times larger than previous surveys of this type. This will be the first longterm deployment for science purposes of a SAPHIRA detector. We have conducted an initial pilot LASSO survey of $\sim$400 stars with Robo-AO installed on the 2.1-m telescope at Kitt Peak, Arizona. The survey will continue in full once Robo-AO is installed on the UH88, 2.1-m telescope, on Maunakea.

\acknowledgments      
Robo-AO KP is a partnership between the California Institute of Technology, the University of Hawai`i, the University of North Carolina at Chapel Hill, the Inter-University Centre for Astronomy and Astrophysics, and the National Central University, Taiwan. Robo-AO KP is supported by a grant from Sudha Murty, Narayan Murthy, and Rohan Murty, and by a grant from the John Templeton Foundation. The Robo-AO instrument was developed with support from the National Science Foundation under grants AST-0906060, AST-0960343, and AST-1207891, the Mt. Cuba Astronomical Foundation, and by a gift from Samuel Oschin.  Based (in part) on observations at Kitt Peak National Observatory, National Optical Astronomy Observatory (NOAO Prop. ID: 15B-3001), which is operated by the Association of Universities for Research in Astronomy (AURA) under cooperative agreement with the National Science Foundation. 


\section*{References}

\begin{enumerate}
\item Atkinson, D., Hall, D. N. B., Baker, I. M., et al., ``Next-generation performance of SAPHIRA HgCdTe APD", High Energy, Optical, and Infrared Detectors for Astronomy VII 9915E, Vol 9915 of Proc. SPIE, pgs 9915-22, 2016.
\item Atkinson, D. et al.,``Dark Current in the SAPHIRA Series of APD Arrays", AJ, Vol. 154, Issue 6, 265, 2017.
\item Baraffe, I. et al., ``New evolutionary models for pre-main sequence and main sequence low-mass stars down to the hydrogen-burning limit", A\&A, 577, A42, 2015.
\item Baranec, C. et al., ``High-efficiency Autonomous Laser Adaptive Optics",ApJ Letters, 790, L8, 2014.
\item Baranec, C., Atkinson, D., Riddle, R., et al., ``High-speed Imaging and Wavefront Sensing with an Infrared Avalanche Photodiode Array", ApJ, 809:70, 2015.
\item Bowler, B., ``Imaging Extrasolar Giant Planets", Publications of the Astronomical Society of the Pacific Journal, Vol 128, Num 968, 2016.
\item Chabrier, G. et al., ``Evolutionary Models for Very Low-Mass Stars and Brown Dwarfs with Dusty Astmospheres", ApJ, 542, 464, 2000.
\item Chabrier, G. et al., ``Giant Planet and Brown Dwarf Formation", Protostars and
Planets VI, Henrik Beuther, Ralf S. Klessen, Cornelis P. Dullemond, and Thomas Henning
(eds.), University of Arizona Press, Tucson, 914 pp., p.619-642, 2014.
\item Eisenhauer, F., Perrin, G. S., Henning, T., et al., Optical and Infrared Interferometry and Imaging V 9907E, Vol 9907 of Proc. SPIE, pgs 9907-6, 2016.
\item Finger, G., Baker, I., Alvarez, D. et al., ``SAPHIRA detector for infrared wavefront sensing", Adaptive Optics Systems IV 9148E, Vol 9148 of Proc. SPIE, pgs 9148-17, 2014.
\item Goebel, S., Guyon, O., Hall, D. N. B., Jovanovic, N., and Atkinson, D., ``Evolutionary timescales of AO-produced speckles at NIR wavelengths", Adaptive Optics Systems V 9909E, Vol 9909 of Proc. SPIE, pgs 9909-46, 2016.
\item Jensen-Clem, R., Duev, D., Riddle, R., Salama, M., et al., ApJ, arXiv:1703.08867, 2017.
\item Muirhead, P. et al., "A Catalog of Cool Dwarf Targets for the Transiting Exoplanet Survey Satellite", 2018.
\item Rodriguez, D. R., Zuckerman, B., Kastner, J. H., et al., ``The GALEX Nearby Young-Star Survey", ApJ, 774, 101, 2013.
\item Salama, M. et al., ``Robo-AO Kitt Peak: status of the system and deployment of a sub-electron readnoise IR camera to detect low-mass companions", Adaptive Optics Systems V 9909E, volume 9909 of Proc. SPIE, pages 99091A-99091A-15., 2016.
\item Schlaufman, K.,``Evidence of an Upper Bound on the Masses of Planets and Its Implications for Giant Planet formation", ApJ, Volume 853, Issue 1, article id. 37, 2018.
\end{enumerate}

  \end{document}